\documentclass[twocolumn,tighten,times]{aastex631}
  
\usepackage[T1]{fontenc}

\newcommand\ergs{erg~s$^{-1}$}

\shorttitle{}
\shortauthors{}

\graphicspath{{./}{fig/}}

\begin{document}

\title{Constraining soft and hard X-ray irradiation in ultraluminous X-ray sources}

\author{Yanli Qiu}
\affiliation{Department of Astronomy, Tsinghua University, Beijing 100084, China}

\correspondingauthor{Hua Feng}
\email{hfeng@tsinghua.edu.cn}

\author[0000-0001-7584-6236]{Hua Feng}
\affiliation{Department of Astronomy, Tsinghua University, Beijing 100084, China}
\affiliation{Department of Engineering Physics, Tsinghua University, Beijing 100084, China}

\begin{abstract}
Most ultraluminous X-ray sources (ULXs) are argued to be powered by supercritical accretion onto compact objects. One of the key questions regarding these objects is whether or not the hard X-rays are geometrically beamed toward the symmetric axis. We propose to test the scenario using disk irradiation, to see how much the outer accretion disk sees the central hard X-rays.  We collect a sample of 11 bright ULXs with an identification of a unique optical counterpart, and model their optical fluxes considering two irradiating sources: soft X-rays from the photosphere of the optically thick wind driven by supercritical accretion, and if needed in addition, hard X-rays from the Comptonization component. Our results indicate that the soft X-ray irradiation can account for the optical emission in the majority of ULXs, and the fraction of hard X-rays reprocessed on the outer disk is constrained to be no more than $\sim$$10^{-2}$ in general. Such an upper limit is well consistent with the irradiation fraction expected in the case of no beaming. Therefore, no stringent constraint on the beaming effect can be placed according to the current data quality.
\end{abstract}

\section{Introduction}

Ultraluminous X-ray sources (ULXs) are off-nuclear point-like X-ray objects with luminosities higher than the Eddington limit of typical stellar mass black holes \citep{Kaaret2017,Fabrika2021}.  It is widely accepted that supercritical accretion occurs in the majority of these sources, especially with the identification of neutron stars in very luminous ULXs \citep[e.g.,][]{Israel2017}.  The physics of supercritical accretion is still unclear.  One of the key questions is whether or not the high apparent luminosity is a result of beaming. 

It is unlikely that most of the ULXs are due to strong beaming or being viewed along a relativistic jet \citep{Koerding2002}.  This scenario contradicts the observed luminosity function \citep{Davis2004}, does not reconcile with the sinusoidal pulse profile observed in pulsar ULXs \citep{Mushtukov2021}, and can be ruled out in cases with surrounding photoionized nebulae \citep{Pakull2002,Kaaret2004, Lehmann2005,Kaaret2009}. 

Geometric beaming may occur due to the presence of a thick accretion flow in supercritical accretion \citep{King2001}.  Recent numerical simulations for supercritical accretion all reveal an optically thin funnel confined by optically thick flows \citep{Jiang2014,Sadowski2016,Takahashi2016,Abarca2018,Kitaki2018,Kitaki2021}. Comptonized X-ray emission from the central region may escape via the funnel and get collimated \citep{Kawashima2012,Sadowski2015,Narayan2017}.   Such a geometry can successfully explain the observed two-component X-ray spectrum, with the soft component interpreted as thermal emission from the outer region (outflow or disk) and the hard component as Comptonized X-rays from the central funnel \citep{Middleton2015}.  Owing to the beaming effect,  the source may appear very soft or supersoft for observers at high inclinations \citep{Feng2016,Soria2016,Urquhart2016,Zhou2019,Qiu2021}.  However,  \citet{Jiang2014} argue that the beaming effect is weak even if X-rays escape from the funnel, because they will undergo multiple scattering and leave the system at the last scattering surface at a large distance, where the collimation is no longer as strong as in the base of the funnel. 

X-ray irradiation on the outer accretion disk can be used for testing the beaming scenario. If there is geometric beaming of the emission in the central funnel, the outer disk will see little or no hard X-rays and consequently no additional optical emission caused by reprocessing of the hard X-rays.  To test this scenario, we adopt the irradiation model under supercritical accretion proposed by \citet{Yao2019}; there are also other models like the one proposed by \citet{Ambrosi2018}. \citet{Yao2019} suggest that supercritical accretion will produce a nearly spherical and optically thick wind \citep{Meier1982a,Zhou2019}. Thus, there are two components of irradiating sources, thermal soft X-rays from the thick wind, and Comptonized hard X-rays from the central funnel.  \citet{Yao2019} found that the former component can well explain the multi-band (X-ray, UV, and optical) spectrum of NGC 247 X-1, which is a supersoft ULX speculated to be viewed at a high inclination angle.  In this model, the soft X-ray irradiation originated from the wind photosphere is always in place, while the hard X-ray irradiation depends on the location and geometry of the last scattering surface. 

In this work, our goal is to investigate whether or not irradiation from hard X-rays or the Comptonization component is needed in addition to the soft component in non-supersoft ULXs. The model is described in \S~\ref{sec:mod}, the sample and data analysis is elaborated in \S~\ref{sec:fit}, and the results are discussed in \S~\ref{sec:dis}. 

\begin{deluxetable*}{lllcllllll}[t]
\tabletypesize{\footnotesize}
\tablecaption{The ULX sample in this work. }
\label{tab:sample}
\tablehead{
\colhead{Name} & \colhead{R.A.} & \colhead{decl.} &  \colhead{$d$}  & \colhead{ebv$_{\rm G}$}  & \colhead{ebv}  &  \colhead{refs.}     &  \colhead{$m$}  &  \colhead{$\dot{m}$} &  \colhead{$L_{\rm C}$} \\
\colhead{} & \colhead{(J2000)} & \colhead{(J2000)} &  \colhead{(Mpc)}  & \colhead{}  & \colhead{}  &  \colhead{}& \colhead{(M$_\odot$)}  &  \colhead{} &  \colhead{($10^{39}$~\ergs)} \\
\colhead{(1)} & \colhead{(2)} & \colhead{(3)}  & \colhead{(4)} & \colhead{(5)} & \colhead{(6)}  & \colhead{(7)} & \colhead{(8)} & \colhead{(9)}  & \colhead{(10)}
}
\startdata
Holmberg II X-1 & 08:19:28.9 & $+$70:42:19 & 3.3 & 0.028  & 0.07 (s) & (1)  & $ 13.2 _{ -4.3 } ^{+ 4.7 } $  &   $ 131.5 _{ -21.2 } ^{+ 19.8 } $    &  $ 10.6  _{ -0.5 } ^{+ 0.6 } $  \\
Holmberg IX X-1 & 09:57:53.2 & $+$69:03:48 & 3.8 & 0.069 & 0.26 (n) & (2)  & $ 19.8 _{ -2.3 } ^{+ 1.9 } $     &    $ 124.4 _{ -9.0 } ^{+ 8.5 } $    &  $ 11.4  _{ -0.2 } ^{+ 0.1 } $    \\
IC 342 X-1 & 03:45:55.6 & $+$68:04:55 & 3.4 & 0.495 & 0.755 (n) &  (3) &$ 17.7 _{ -9.1 } ^{+ 12.5 } $     &    $ 131.4 _{ -47.1 } ^{+ 30.7 } $    &  $ 4.6  _{ -0.5 } ^{+ 1.1 } $     \\
NGC 55 ULX1 & 00:15:28.9 & $-$39:13:18 & 1.8 & 0.012 & 0.41 (x)  & (4)  &  $ 15.7 _{ -0.7 } ^{+ 1.2 } $     &    $ 157.3 _{ -8.9 } ^{+ 5.3 } $    &  $ 0.39  _{ -0.01 } ^{+ 0.02 } $  \\
NGC 1313 X-1 & 03:18:20.0 & $-$66:29:11 & 4.3  & 0.097 & 0.82 (x) & (4)   &$ 41.3 _{ -3.4 } ^{+ 3.5 } $     &    $ 102.0 _{ -8.3 } ^{+ 8.2 } $    &  $ 9.7  _{ -0.4 } ^{+ 0.5 } $   \\
NGC 1313 X-2 & 03:18:22.1 & $-$66:36:03 & 4.3 & 0.075 & 0.13 (n) & (5)  &  1.4 &    $ 186.9 _{ -14.1 } ^{+ 15.1 } $    &  $ 1.9  _{ -0.1 } ^{+ 0.1 } $  \\
NGC 2403 X-1 & 07:36:25.5 & $+$65:35:39 & 4.2 & 0.036 & 0.27 (x)  & (4)  & $ 19.7 _{ -3.8 } ^{+ 36.6 } $     &    $ 85.4 _{ -50.1 } ^{+ 16.2 } $    &  $ 2.4  _{ -0.1 } ^{+ 0.1 } $    \\
NGC 4559 ULX1 & 12:35:51.7 & $+$27:56:04 & 9.7 & 0.016 & 0.10 (n) & (6)  &  $ 71.0 _{ -10.8 } ^{+ 25.8 } $     &    $ 115.9 _{ -23.6 } ^{+ 18.6 } $    &  $ 8.0  _{ -0.5 } ^{+ 2.3 } $  \\
NGC 5204 X-1 & 13:29:38.6 & $+$58:25:06 & 4.9 & 0.011 & 0.11 (n) & (6)   &   $ 7.3 _{ -1.6 } ^{+ 1.6 } $     &    $ 130.6 _{ -17.1 } ^{+ 21.7 } $    &  $ 5.1  _{ -0.1 } ^{+ 0.2 } $   \\
NGC 5408 X-1 & 14:03:19.6 & $-$41:22:58 &  4.8 & 0.061 & 0.08 (n) & (7)  &  $ 31.8 _{ -1.6 } ^{+ 1.5 } $     &    $ 137.2 _{ -5.0 } ^{+ 4.6 } $    &  $ 5.9  _{ -0.3 } ^{+ 0.5 } $    \\
NGC 6946 X-1 & 20:35:00.7 & $+$60:11:30 & 7.7 & 0.302 & 0.50 (n) & (8)  &   $ 102.8 _{ -12.0 } ^{+ 15.6 } $     &    $ 93.5 _{ -8.1 } ^{+ 7.7 } $    &  $ 5.6  _{ -0.2 } ^{+ 0.3 } $    \\
\enddata
\tablecomments{
Col.~1: source name.
Col.~2: right ascension from 4XMM-DR9 \citep{Webb2020}.
Col.~3: declination from 4XMM-DR9.
Col.~4: distance to the host galaxy \citep[see][for the references]{Qiu2021}.
Col.~5: Galactic $E(B-V)$ along the line of sight to the source \citep{Schlafly2011}.
Col.~6: Total $E(B-V)$, which is derived from surrounding nebula (n), nearby stellar population (s), or converted from X-ray absorption (x).
Col.~7: references for Col.~6.
Col.~8: dimensionless compact object mass, $m = M / M_\sun$, estimated from X-ray modeling \citep{Qiu2021}.
Col.~9: dimensionless mass accretion rate, $\dot{m} = \dot{M}/\dot{M}_{\rm Edd}$, also from X-ray modeling \citep{Qiu2021}, where $\dot{M}_{\rm Edd} = L_{\rm Edd}/0.1c^2$.
Col.~10: the 0.3--10 keV isotropic X-ray luminosity of the Comptonization component \citep{Qiu2021}.
}
\tablerefs{(1) \citet{Stewart2000}, (2) \citet{Grise2011}, (3) \citet{Grise2006}, (4) \citet{Qiu2021}, (5) \citet{Grise2008}, (6) \citet{Vinokurov2018}, (7) \citet{Kaaret2009}, (8) \citet{Abolmasov2008}.
}
\end{deluxetable*}

\section{The irradiation Model}
\label{sec:mod}

The model proposed by \citet{Yao2019} is based on a geometry that incorporates an optically thick wind on top of a slim disk \citep{Meier1982a,Meier2012,Zhou2019}.  The wind originates from the advective radius of the slim disk, where the luminosity approaches the Eddington limit and thus massive winds are launched.  This is the wind base and the disk properties at this radius are assumed to be the boundary conditions of the wind. The wind develops outwards following radiation hydrodynamic solutions. There are two characteristic radii, the photospheric radius where the gas and radiation are in equilibrium at the photospheric temperature, and the radius of the last scattering surface (or the so-called scattersphere) where the photons move outwards freely.  The wind is optically thin for absorption but optically thick for scattering between the two radii.  The disk within the two radii is heated by the local radiative flux in the wind directly, and the disk beyond the scattersphere is heated by the emission emergent on the scattersphere.  The soft irradiation from the wind is determined by two parameters, the mass of compact object, $m$, and the mass accretion rate, $\dot{m}$.  Here, $m$ and $\dot{m}$ are dimensionless parameters, normalized to the solar mass and the critical accretion rate, respectively. The critical accretion rate is defined as the rate just needed to power the Eddington limit ($L_{\rm Edd}$ ) assuming an efficiency of 0.1, i.e., $\dot{M}_{\rm crit} = 10L_{\rm Edd} / c^2$.  In this paper, the term ``supercritical accretion'' refers to the accretion rate. This model provides an adequate fit to the X-ray/UV/optical emission of NGC 247 X-1 \citep{Yao2019}, and can also successfully explain the cool thermal emission seen in the energy spectra of luminous and very soft X-ray sources \citep{Zhou2019} and ULXs~\citep{Qiu2021}.  

Even with high accretion rates, the standard accretion model \citep{Shakura1973} is a good approximation for the outer disk, e.g., beyond the spherization radius.  For irradiation due to hard X-rays, a simplified assumption would be a point-like X-ray source at the center of the system shining at a standard outer disk \citep[e.g.,][]{Gierlinski2009}.  This may not be valid in the case of supercritical accretion, as the hard X-rays are supposed to originate from the central funnel, where the scattering optical depth is high, and thus the X-rays appear to leave from the scattersphere of the funnel at a high altitude.  The height of the scattersphere and the angular distribution of photons on the scattersphere determines the geometry of irradiation. To avoid these uncertainties, here we give up the detailed geometry but adopt a simplified factor, $f_{\rm out}$, to describe the fraction of hard X-rays from the Comptonization component reprocessed on the outer disk.  In principle, $f_{\rm out}$ is a function of radius due to different interception angles, but here we assume it is a constant throughout the disk; or it can be understood as the average value.

Thus, the hard irradiation at different radii can be written as $T_{\rm irr, hard}^4(r) = f_{\rm out} L_{\rm C} / (4 \pi \sigma r^2 )$, where $r$ is the radius, $L_{\rm C}$ is the apparent luminosity of the Comptonization component, and $\sigma$ is the Stefan-Boltzmann constant.  We note that $f_{\rm out}$ is defined against the apparent luminosity instead of the unknown, bolometric luminosity. Thus,  $f_{\rm out}$ does not refer to the actual fraction, but $f_{\rm out} L_{\rm C}$ does point to the absolute reprocessed luminosity.

Together with the intrinsic (viscous) temperature of the standard disk, $T_{\rm vis}$, and the soft irradiation temperature $T_{\rm irr, soft}$ \citep[see][for details of the implementation]{Yao2019}, the outer disk temperature is obtained as $T_{\rm eff}^4 (r) = T_{\rm vis}^4 (r) + T_{\rm irr,soft}^4(r) + T_{\rm irr,hard}^4(r)$, and $T_{\rm vis}$ can be ignored almost in all cases. Integration of the blackbody spectra over the radius all the way to the outermost disk radius, $r_{\rm out}$, gives the irradiation spectrum. The soft irradiation is integrated from the wind photosphere to  $r_{\rm out}$. The hard irradiation is not sensitive to the inner radius of the integration, because the majority of the intercepted luminosity is deposited on the outermost part of the disk. We thus set the innermost radius as the wind scattersphere radius in the integration, because external emission into the scattersphere will be significantly diluted and attenuated when it reaches the disk.

For the soft irradiation, we assume a disk albedo $\beta=0.7$ and a viscosity parameter $\alpha=0.1$ \citep{Yao2019}. For the hard irradiation, there is no need to assume an albedo as it is absorbed in $f_{\rm out}$.  In sum, there are 6 parameters in the irradiation model: $m$, $\dot{m}$, $L_{\rm C}$, $r_{\rm out}$, $f_{\rm out}$, and the disk inclination $i$.  In this model, the cool blackbody component in the X-ray band is assumed to be the thermal emission originated from the wind photosphere, from which $m$ and $\dot{m}$ can be determined via X-ray spectral fitting. $L_{\rm C}$ is also derived from X-ray fitting with a Comptonization component. Other parameters are constrained by fitting with the optical spectral energy distribution (SED).  

\section{Sample and SED fit}
\label{sec:fit}

The sample is based on the bright ULXs studied in \citet{Qiu2021}.  We require an identification of a unique optical counterpart for each ULX, and thus have to discard M51 ULX7 and ULX8 \citep{Terashima2006,Earnshaw2016}. NGC 300 ULX1 is not included because its optical emission is dominated by the donor star \citep{Heida2019}. The final sample contains 11 bright ULXs, whose optical emission is argued to be dominated by disk irradiation \citep{Tao2011,Gladstone2013}, and is listed in Table~\ref{tab:sample}.  

The X-ray spectra of these ULXs have been analyzed in \citet{Qiu2021}, using a model including a blackbody component, to account for thermal emission from the photosphere of the optically thick wind, and a Comptonization component, for hard X-rays from the central funnel. The parameters that are useful for this work include the blackbody temperature and luminosity, which are associated with $m$ and $\dot{m}$ in the wind model \citep{Meier2012,Zhou2019}, and the Comptonization luminosity $L_{\rm C}$. For the pulsar ULX NGC 1313 X-2 \citep{Sathyaprakash2019},  the compact object mass is fixed at 1.4~$M_\sun$ in the X-ray fitting.  If multiple observations are available for one source, we adopt the one having the largest number of detected photons.  We find that the fitting results are consistent with each other with different combinations of $m$, $\dot{m}$, and $L_{\rm C}$ derived from different observations, validating the choice of the observation with the best statistics.

The optical fluxes are adopted from the literature \citep{Yang2011,Tao2011,Gladstone2013}. The Galactic extinction along the line of sight to each object is adopted from \citet{Schlafly2011}, set as the lower bound. The total extinction is quoted from the literature, in which it is derived from the surrounding ULX nebula using Balmer decrement or by modeling the nearby stellar population; otherwise, it is converted from the X-ray absorption column density $N_{\rm H}$ quoted from \citet{Qiu2021} following the Milky Way relationship $N_{\rm H} / A_V= (2.21 \pm 0.09) \times 10^{21}$~cm$^{-2}$~mag$^{-1}$~\citep{Guever2009}. If there are multiple optical observations available in one band, we find that they have negligible variation and thus adopt the one with the median flux. 

\begin{figure*}[t]
\includegraphics[width=0.33\textwidth]{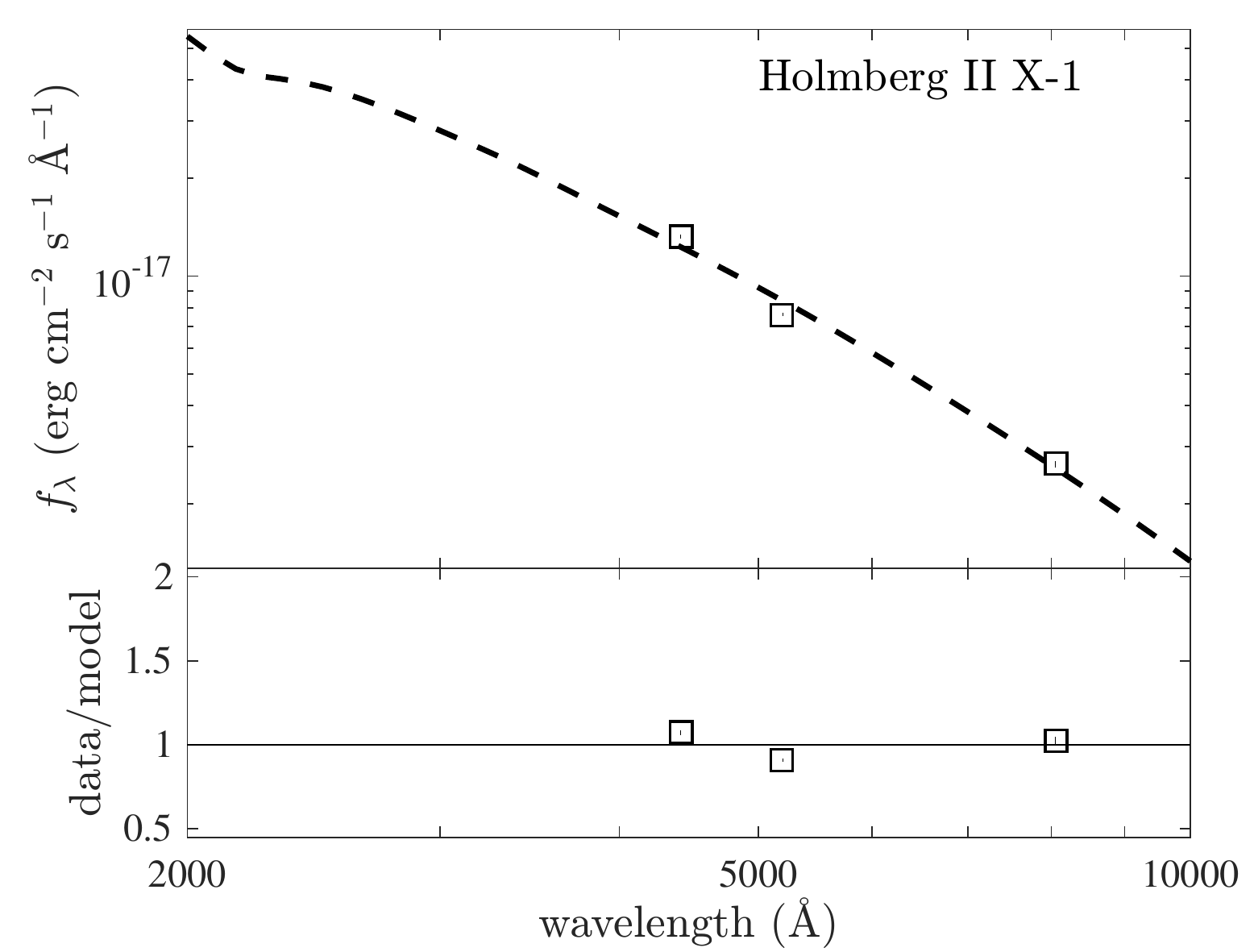}
\includegraphics[width=0.33\textwidth]{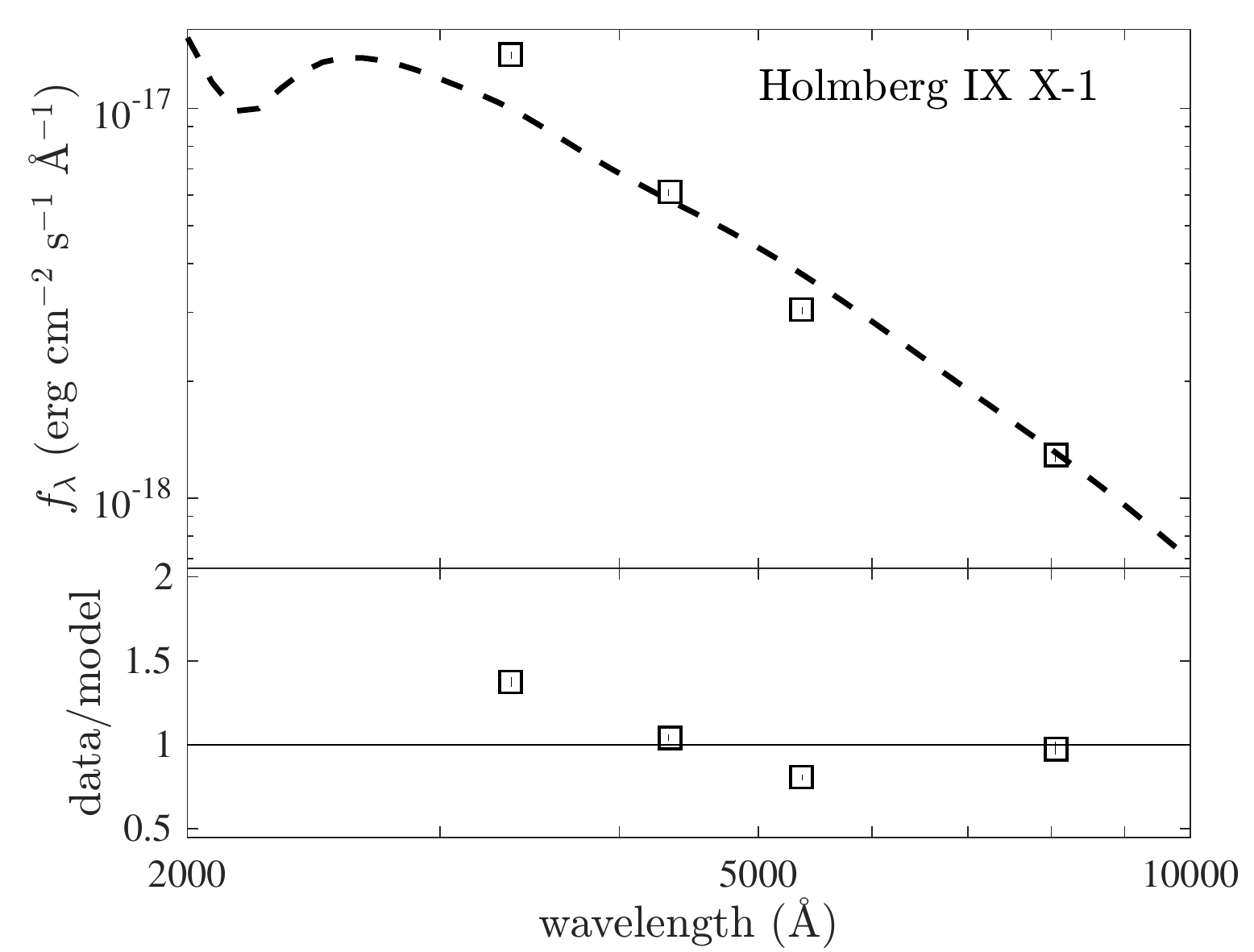}
\includegraphics[width=0.33\textwidth]{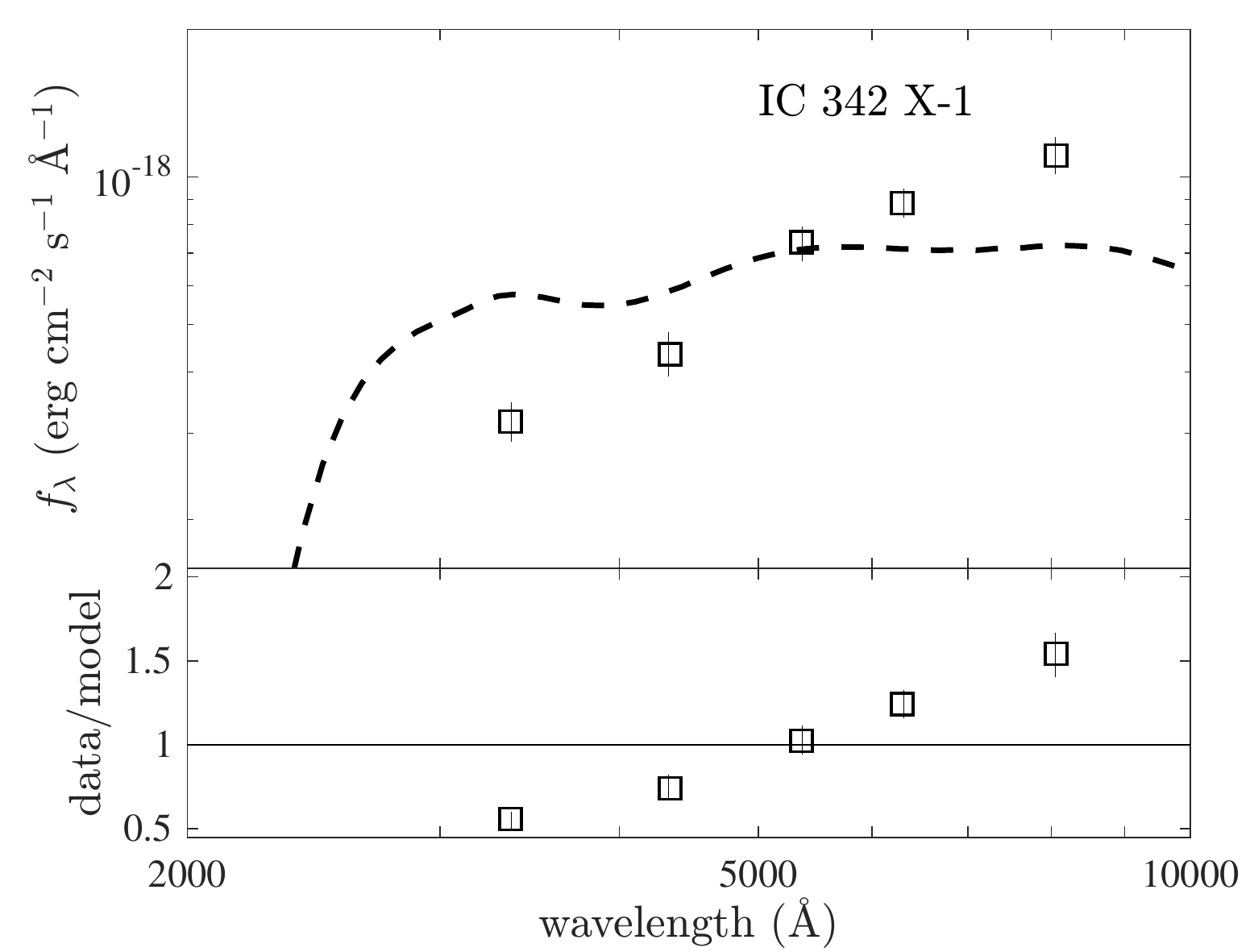}
\includegraphics[width=0.33\textwidth]{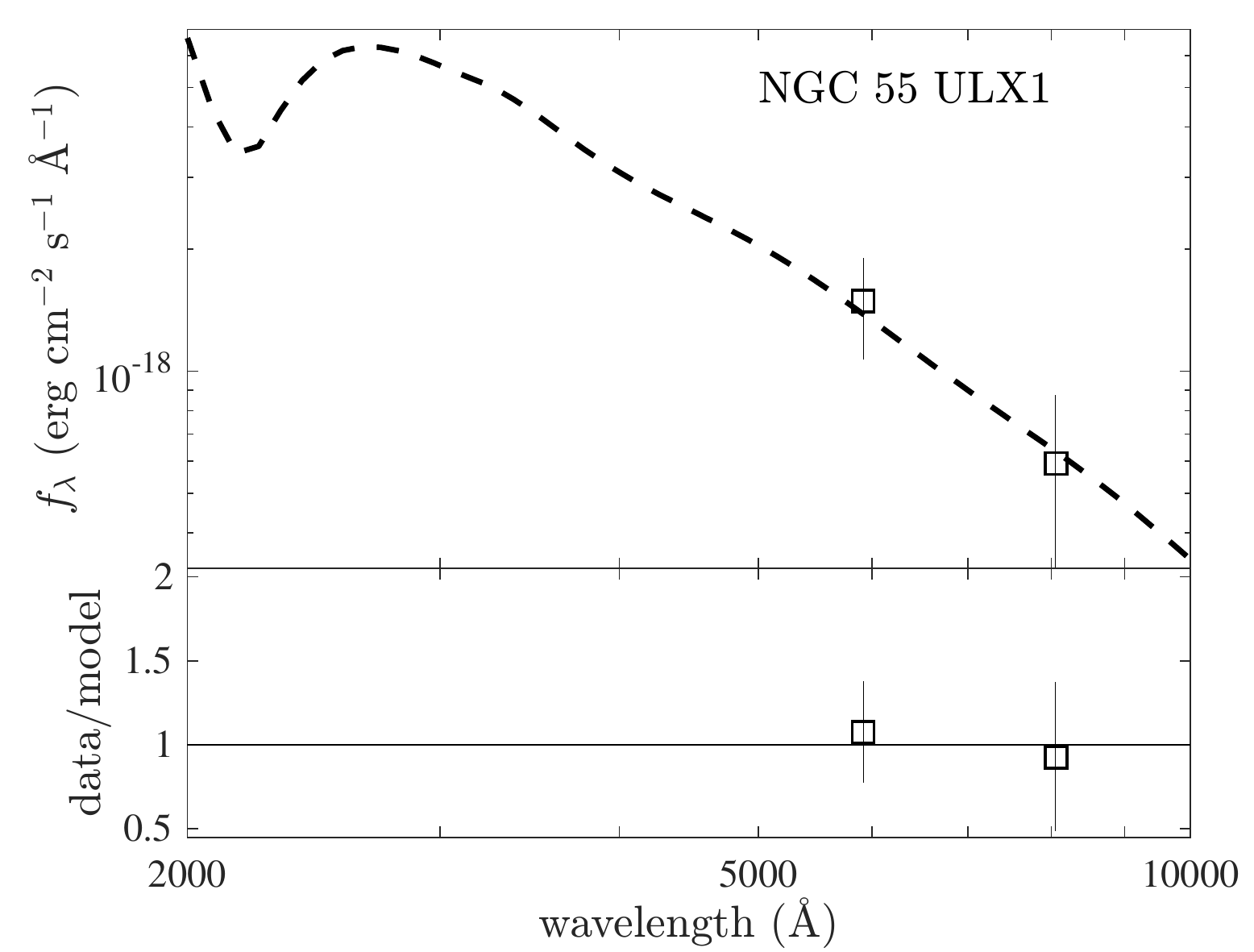}
\includegraphics[width=0.33\textwidth]{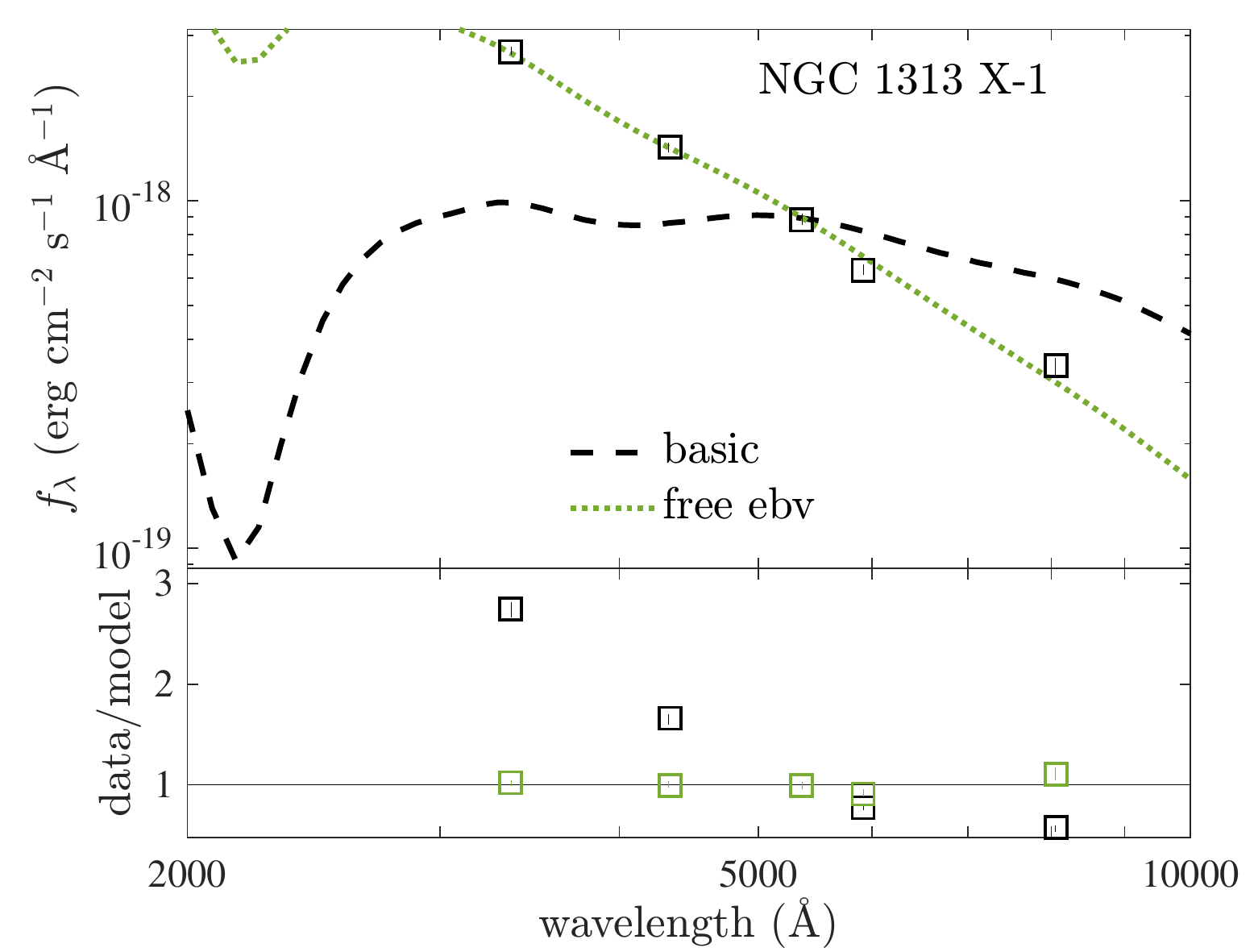}
\includegraphics[width=0.33\textwidth]{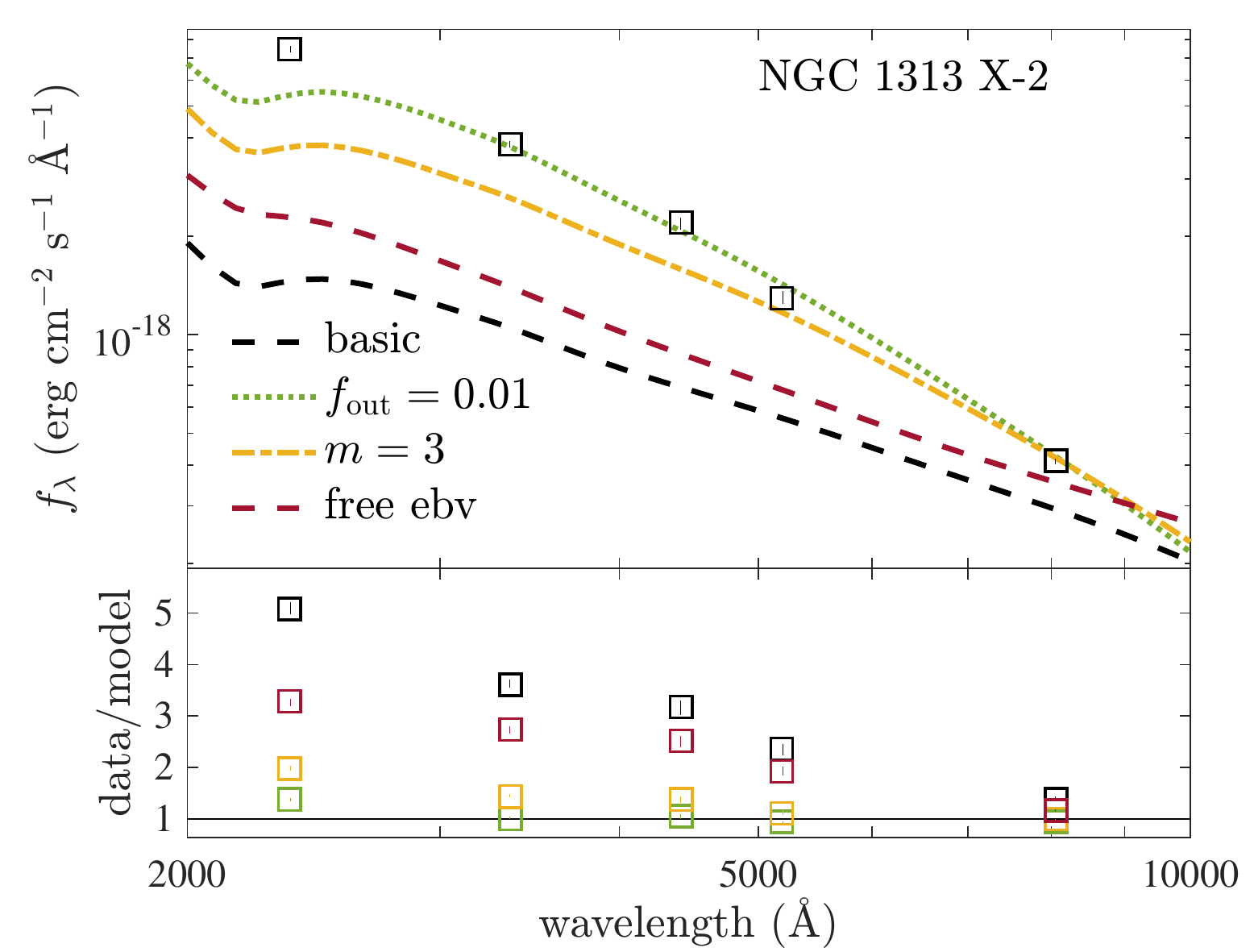}
\includegraphics[width=0.33\textwidth]{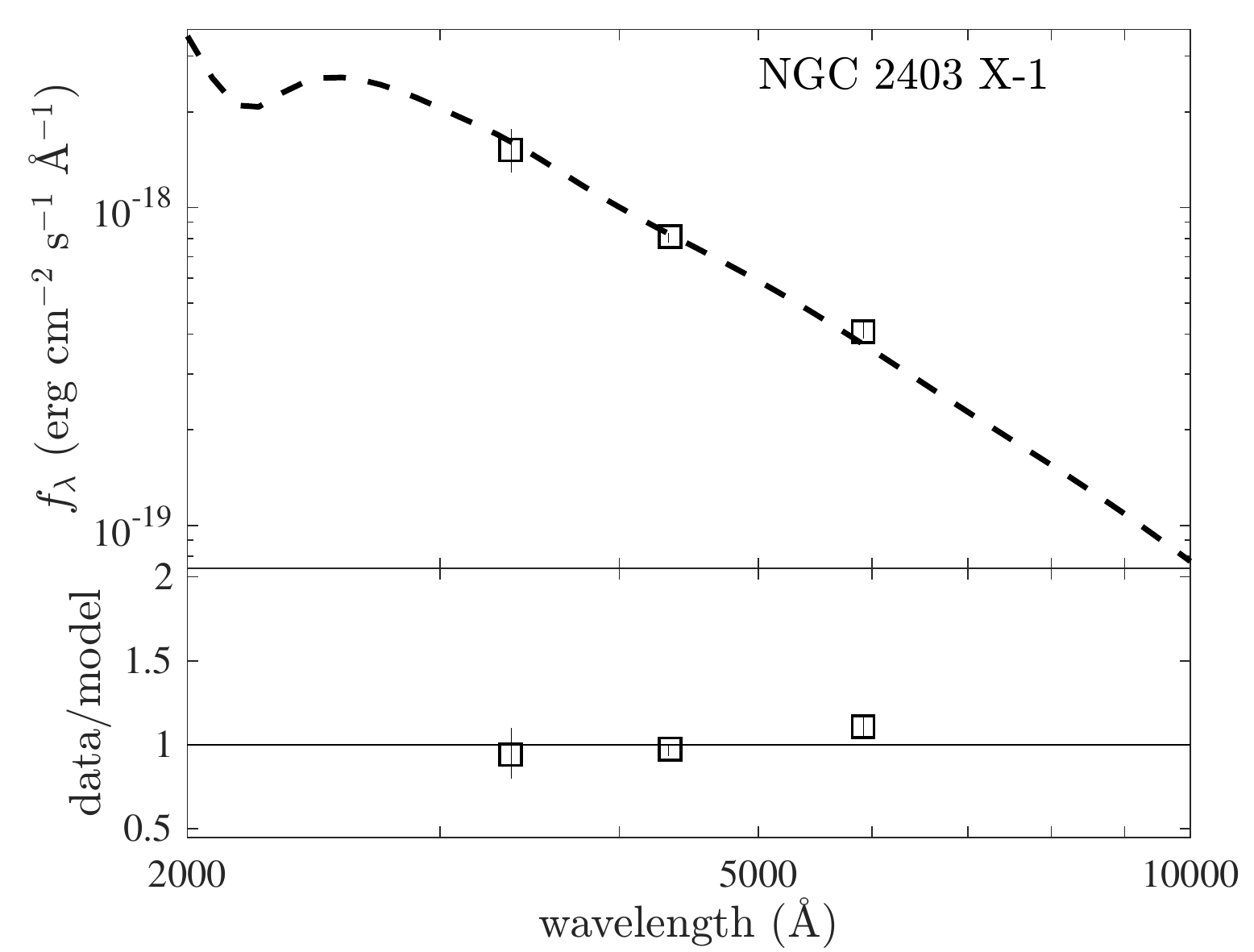}
\includegraphics[width=0.33\textwidth]{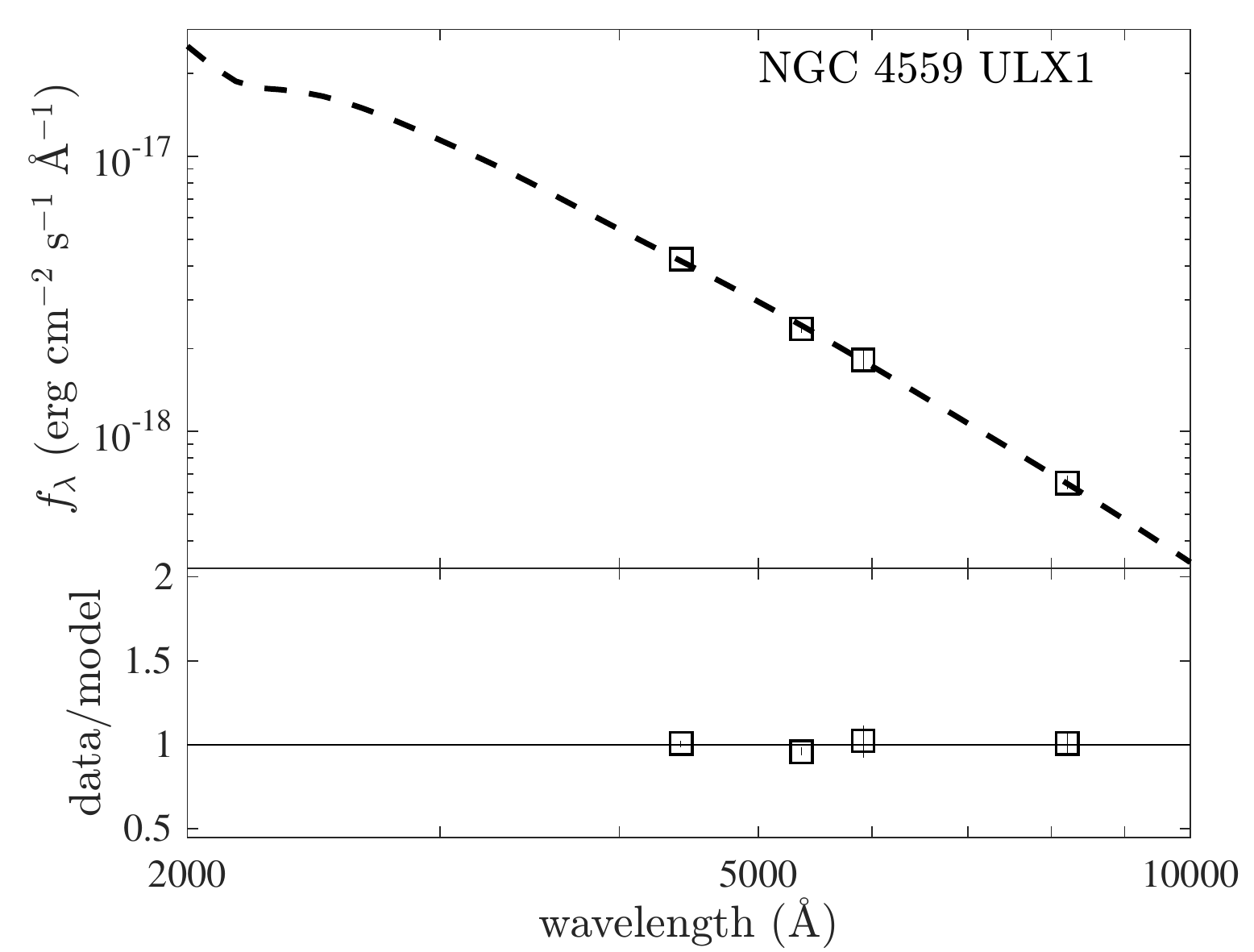}
\includegraphics[width=0.33\textwidth]{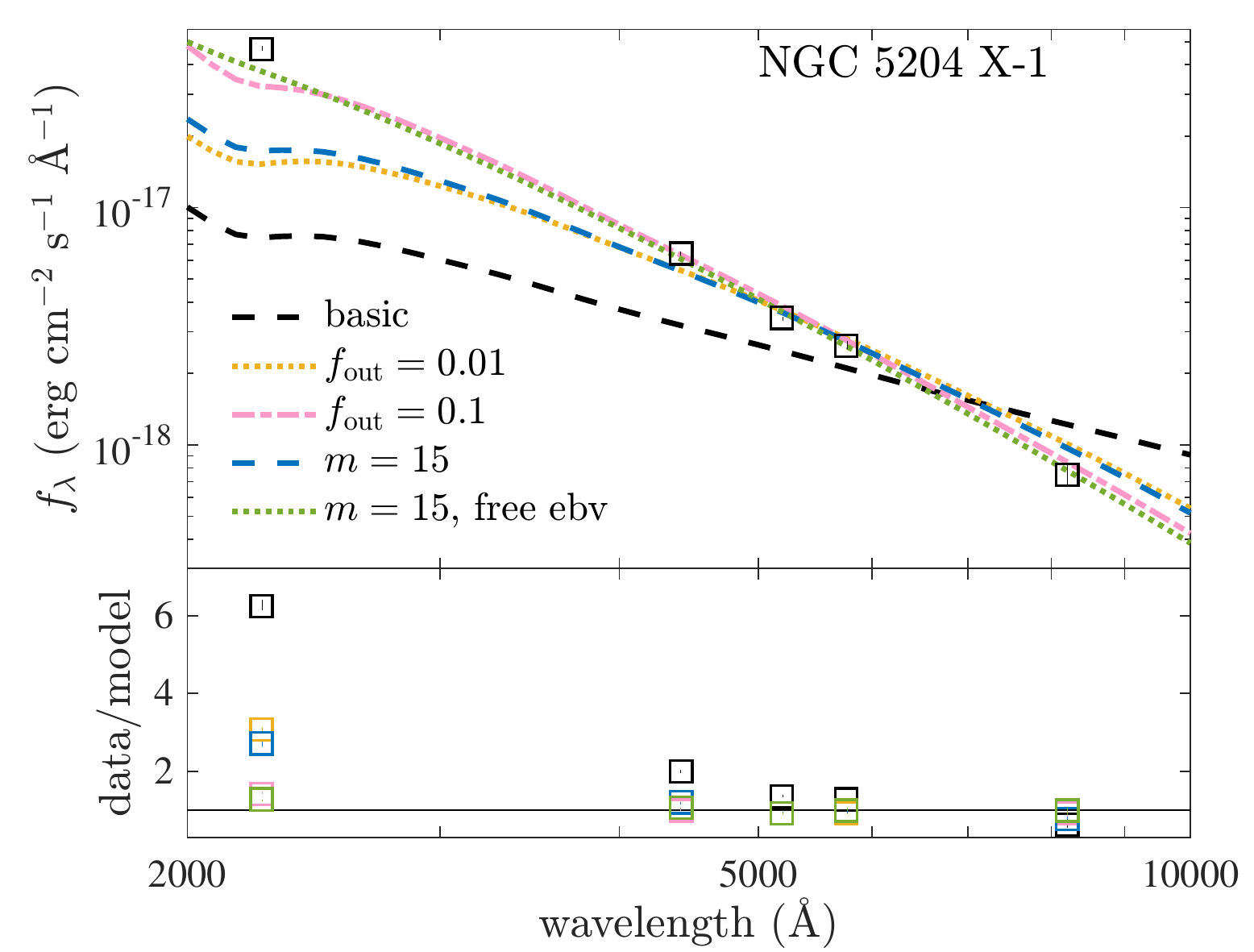}
\includegraphics[width=0.33\textwidth]{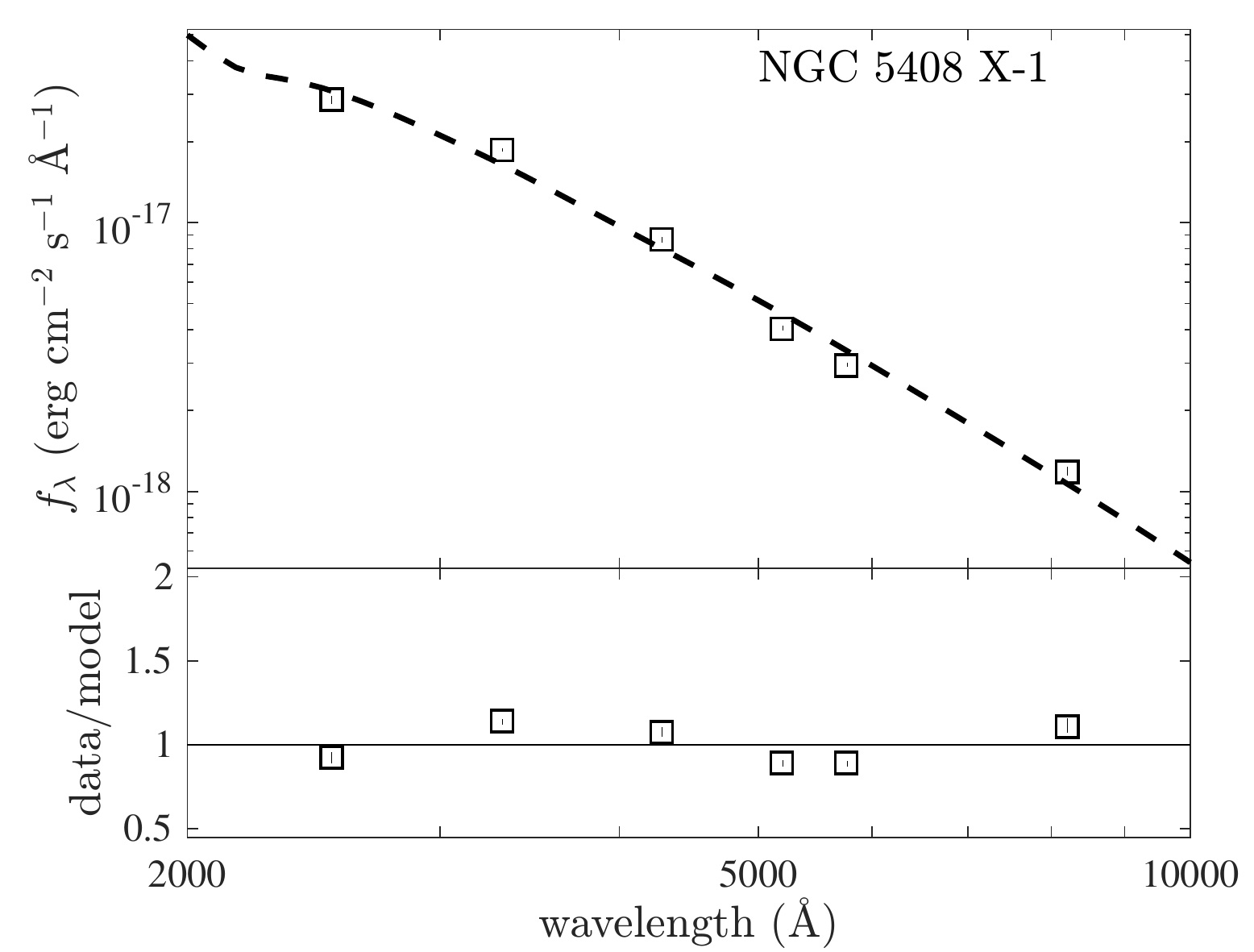}
\includegraphics[width=0.33\textwidth]{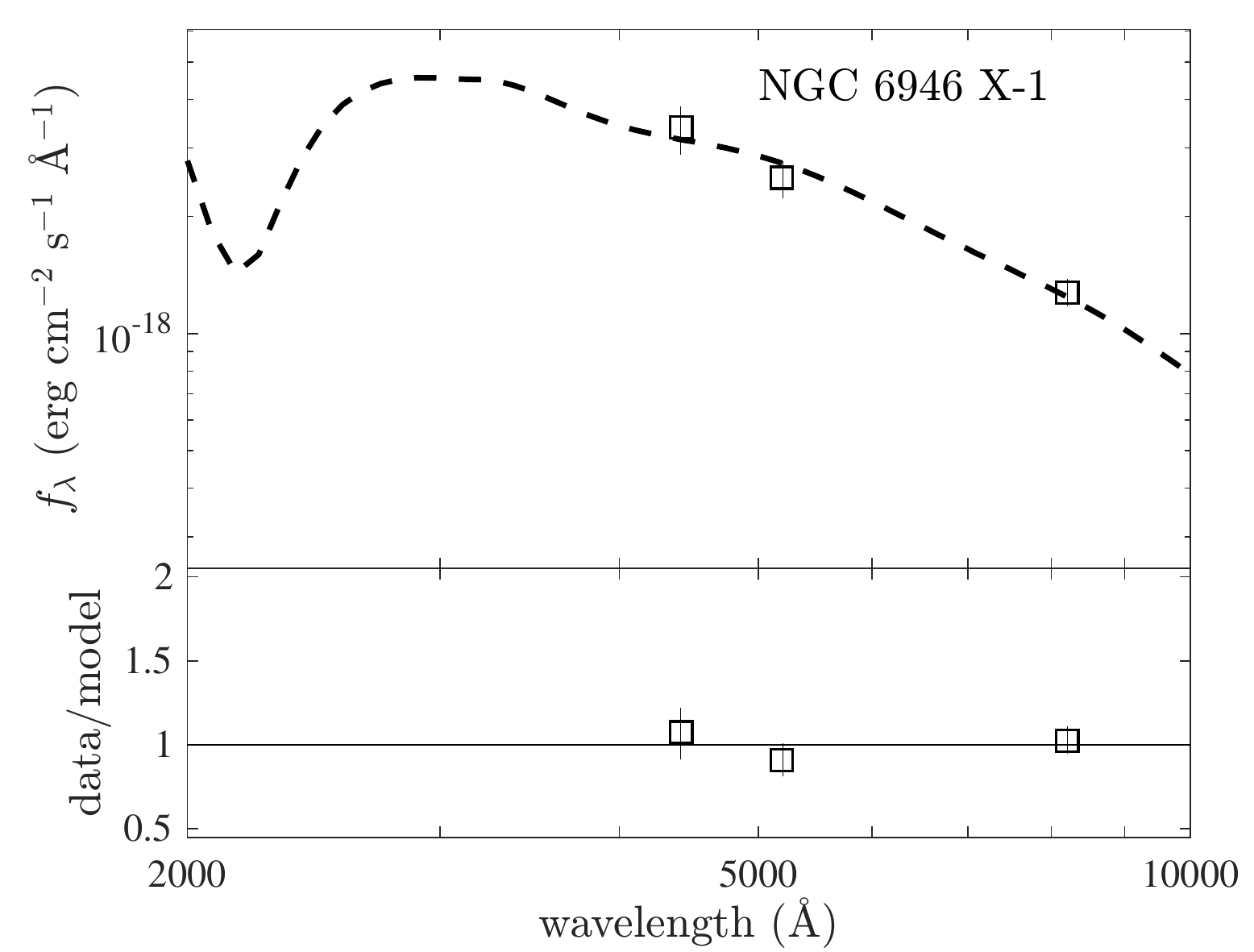}
\caption{Optical SEDs of the ULXs and best-fit irradiation models. The black dashed line indicates the basic model that assumes $f_{\rm out}=0$ and parameters ($m$, $\dot{m}$, $L_{\rm C}$, and $E(B-V)$) with values in Table~\ref{tab:sample}. Addition model assumptions are applied for the three sources (NGC 1313 X-1, X-2, and NGC 5204 X-1) that the basic model fails to fit the data. See the text for details about the model parameters.}
\label{fig:fit}
\end{figure*}

First, we fit the optical SED with the basic model, which assumes $f_{\rm out} = 0$, i.e., there is no hard irradiation, and fix $m$, $\dot{m}$, $L_{\rm C}$, and $E(B-V)$ at values quoted in Table~\ref{tab:sample}. This leaves the outermost disk radius $r_{\rm out}$ and the cosine inclination $\cos i$ as two free parameters. $r_{\rm out}$ is searched from the scattersphere radius to $10^{15}$~cm in the logarithmic space. $\cos i$ is searched from 0 to 1.  Due to possible uncertainties or systematics in the distance, extinction, and model simplifications, it is not meaningful to minimize the $\chi^2$.  We thus assume that the optical measurement errors are much less than the unknown systematic error, and find the best-fit parameters using a least-squares minimization on the logarithmic flux.  In this case, the errors on the parameters reflect the deviation between the model and data.  The fitting results are listed in Table~\ref{tab:fit} and shown in Figure~\ref{fig:fit}.  We consider the fit acceptable if the data to model flux ratios are all within 1/2 and 2, which is achieved for 8 out of the 11 sources.  

For sources with an acceptable fit with soft irradiation, we calculate the reprocessed power on the outer disk, and compare it with the Comptonization luminosity. The ratios are mostly on the order of $10^{-2}$ and have a median of 0.02. NGC 55 ULX1 is the only outlier with a ratio as high as 0.5, because the source is relatively soft and has a relatively low Comptonization luminosity.  As the reprocessed optical flux is scaled with the irradiation power, this indicates that if there is additional hard irradiation with $f_{\rm out} \approx 10^{-2}$, the data to model flux ratio will vary by a factor of 2, i.e., from $\sim$1 (see Table~\ref{tab:fit}) to nearly 0.5, which is still considered acceptable.  Therefore, we argue that, for these 8 sources in which hard X-ray irradiation is not needed in the fitting, we place an upper limit of $\sim$$10^{-2}$ on the hard irradiation fraction $f_{\rm out}$.

\begin{deluxetable}{lccc}
\tabletypesize{\footnotesize}
\tablecaption{Fitting results with the basic model that assumes $f_{\rm out} = 0$ and the parameters $m$, $\dot{m}$, $L_{\rm C}$, and $E(B-V)$ with values quoted in Table~\ref{tab:sample}.}
\label{tab:fit}
\tablehead{
\colhead{Name} & \colhead{$\cos i$} & \colhead{$\log(r_{\rm out}/{\rm cm})$} & \colhead{$R_{\min}/R_{\max}$}\\
\colhead{(1)} & \colhead{(2)} & \colhead{(3)} & \colhead{(4)}
}
\startdata
Holmberg II X-1  &   $ 0.82 _{- 0.55 } ^{+ 0.18 } $  &    $ 12.5 _{- 0.8 } ^{+ 0.8 } $    & 0.9 / 1.1    \\
Holmberg IX X-1  &   $ 1.00 _{- 1.00 } ^{+ 0.00 } $  &    $ 12.3 _{- 0.6 } ^{+ 1.3 } $    & 0.8 / 1.4    \\
IC 342 X-1  &   $ 0.29 _{- 0.29 } ^{+ 0.39 } $  &    $ 13.2 _{- 1.6 } ^{+ 1.8 } $    & 0.6 / 1.5    \\
NGC 55 ULX1  &  0.31 &    11.8   &  0.9 / 1.1    \\
NGC 1313 X-1 (x)  &   $ 0.97 _{- 0.97 } ^{+ 0.03 } $  &    $ 12.3 _{- 0.5 } ^{+ 2.7 } $    & 0.6 / 2.7    \\
NGC 1313 X-2  &   $ 1.00 _{- 1.00 } ^{+ 0.00 } $  &    $ 12.7 _{- 2.0 } ^{+ 2.3 } $    & 1.4 / 5.1    \\
NGC 2403 X-1  &   $ 0.37 _{- 0.37 } ^{+ 0.63 } $  &    $ 11.9 _{- 0.0 } ^{+ 0.0 } $    & 0.9 / 1.1    \\
NGC 4559 ULX1  &   $ 0.38 _{- 0.13 } ^{+ 0.13 } $  &    $ 12.7 _{- 0.0 } ^{+ 0.0 } $    & 1.0 / 1.0    \\
NGC 5204 X-1  &   $ 1.00 _{- 1.00 } ^{+ 0.00 } $  &    $ 13.6 _{- 2.4 } ^{+ 1.4 } $    & 0.6 / 6.3    \\
NGC 5408 X-1  &   $ 0.42 _{- 0.42 } ^{+ 0.58 } $  &    $ 12.4 _{- 0.5 } ^{+ 0.8 } $    & 0.9 / 1.1    \\
NGC 6946 X-1  &   $ 0.71 _{- 0.71 } ^{+ 0.29 } $  &    $ 12.8 _{- 0.6 } ^{+ 0.7 } $    & 0.9 / 1.1    \\
\enddata
\tablecomments{
Col.~1: source name.
Col.~2: the cosine inclination.
Col.~3: the logarithmic disk outer radius; the error is shown as 0.0 due to rounding off.
Col.~4: the minimum/maximum data to model ratio.  
The errors for NGC 55 ULX1 are not quoted because there is no degree of freedom.
}
\end{deluxetable}

\begin{deluxetable*}{lcccccc}
\tabletypesize{\footnotesize}
\tablecaption{Addition fittings for the three exceptional sources assuming different $f_{\rm out}$ or $m$, and/or a relaxation of the extinction.}
\label{tab:fit3}
\tablehead{
\colhead{Name} &\colhead{$f_{\rm out}$} & \colhead{$m$} & \colhead{$E(B-V)$} & \colhead{$\cos i$} & \colhead{$\log(r_{\rm out}/{\rm cm})$} & \colhead{$R_{\min}/R_{\max}$}\\
\colhead{(1)} & \colhead{(2)} & \colhead{(3)} & \colhead{(4)} & \colhead{(5)} & \colhead{(6)} & \colhead{(7)}
}
\startdata
NGC 1313 X-1  & 0 &  41.3 & $0.35_{-0.26}^{+0.46}$ & $ 0.82 _{- 0.82 } ^{+ 0.18 } $  &    $ 11.9 _{- 0.2 } ^{+ 0.1 } $    &   0.9/1.1   \\ 
\noalign{\smallskip}\hline\noalign{\smallskip}   
NGC 1313 X-2 & $ 0.01$  &  1.4 & 0.13 &   $1.00_{-1.00}^{+0.00}$    &    $12.1_{-0.2}^{+1.1}$ &  0.9/1.4   \\ 
NGC 1313 X-2 & $ 0$  &  3 & 0.13 &   $1.00_{-1.00}^{+0.00}$  &   $12.3_{-1.5}^{+2.7}$    &  1.0/2.0    \\ 
NGC 1313 X-2 & $ 0$  &  1.4 & $0.08_{-0.00}^{+0.18}$ &   $1.00_{-1.00}^{+0.00}$    &    $15.0_{-4.3}^{+0.0}$ &  1.2/3.3   \\ 
\noalign{\smallskip}\hline\noalign{\smallskip}   
NGC 5204 X-1 & 0.01  & 7.3 & 0.11 & $1.00_{-0.91}^{+0.00}$    &  $12.4_{-1.1}^{+2.6}$     &  0.8/3.1\\ 
NGC 5204 X-1 & 0.1  & 7.3 & 0.11 &  $ 1.00 _{- 1.00 } ^{+ 0.00 } $   &    $12.1_{-0.8}^{+1.9}$  &    0.9/1.4 \\ 
NGC 5204 X-1 & 0  & 15 & 0.11 &  $1.00_{-1.00}^{+0.00}$   &   $12.3_{-1.1}^{+2.7}$  &  0.78/2.7   \\ 
NGC 5204 X-1 & 0  & 15 & $0.01_{-0.00}^{+0.03}$ &   $1.00_{-1.00}^{+0.00}$ &   $12.2_{-0.9}^{+2.8}$ & 0.94/1.25    \\
\enddata
\tablecomments{
Col.~1: source name.
Col.~2: fraction of the Comptonization luminosity reprocessed on the outer disk.
Col.~3: dimensionless compact object mass.
Col.~4: total extinction along the line of sight.
Col.~5: the cosine inclination.
Col.~6: the logarithmic disk outer radius.
Col.~7: the minimum/maximum data to model ratio.  
Parameters without errors are fixed during the fit.
}
\end{deluxetable*}

\subsection{Three exceptions}

NGC 1313 X-1, X-2, and NGC 5204 X-1 are the three sources without an acceptable fit with the basic model.  For NGC 1313 X-1, the X-ray-converted extinction could be overestimated, as the X-ray emitting region is more compact and possibly having more materials for attenuation than in optical. Therefore, we set the extinction as a free parameter and fit the SED again, leading to an acceptable fit with an $E(B-V)$ of 0.35, a number between the Galactic and X-ray converted values.  

For NGC 1313 X-2 and NGC 5204 X-1, we regard the optical extinction reliable to some extent because it is derived from the surrounding nebula on the basis of the Balmer decrement.  We thus try to improve the fit assuming $f_{\rm out}  = 0.01$ or 0.1.  $f_{\rm out} = 0.01$ can significantly improve the fit for NGC 1313 X-2, and $f_{\rm out} = 0.1$ works for NGC 5204 X-1.  

The second attempt is to keep $f_{\rm out}  = 0$ but assume a higher compact object mass to recover the model deficit.  We note that the model flux is not much sensitive to $\dot{m}$ because the soft X-ray luminosity from the wind photosphere is limited at Eddington, which is determined only by $m$.  For NGC 1313 X-2, which is a pulsar ULX \citep{Sathyaprakash2019}, we therefore assume $m = 3$ to represent the possibly most massive neutron star in theory \citep{Lattimer2012}, and obtain a marginally acceptable fit. For NGC 5204 X-1, we have $m = 7.3$ in the basic model. We examine the X-ray fitting and find that acceptable fits can be obtained only with $m < 15$. We thus assume $m = 15$ for a massive black hole, but the fit is still not acceptable. 

Although the extinction for these two sources is estimated in a way more reliable than others, we caution that the extinction is found to vary within the nebula region \citep{Grise2011}, and it is always challenging to find the true value to a point-like object. However, simply varying the extinction as per the basic model cannot produce an acceptable fit for both sources.  Thus, as the third attempt, we fit the SED assuming $m = 3$ with a free extinction for NGC 1313 X-2, leading to a significant improvement.  For NGC 5204 X-1, $m = 15$ plus a free extinction also works. 

To conclude,  NGC 1313 X-1 can be well fitted with a reasonable extinction. For NGC 1313 X-2, there are two ways to obtain an acceptable fit, either by assuming a massive neutron star (even better fitted with a Galactic extinction) or introducing some hard irradiation with $f_{\rm out} = 0.01$. For NGC 5204 X-1, either a massive black hole with Galactic extinction or $f_{\rm out} = 0.1$ can fit the SED.  The above fitting results for the three sources are listed in Table~\ref{tab:fit3} and displayed in Figure~\ref{fig:fit}.
 
\section{Discussion}
\label{sec:dis}

In this work, we investigate the disk irradiation in non-supersoft ULXs, to see if there is hard X-ray (from the Comptonization component) irradiation in addition to the soft X-ray (from the wind photosphere) irradiation. This may help answering an important question in the context of supercritical accretion: is there any beaming effect for the hard X-ray emission? We collect a sample of 11 ULXs, and test the scenario based on the irradiation model proposed by \citet{Yao2019} that involves radiative transfer in an optically thick radiation-driven wind.  The degree of hard irradiation is specified with the parameter $f_{\rm out}$, which determines the fraction of apparent Comptonization luminosity being reprocessed on the outer accretion disk.  To conclude, for the majority of sources, the soft irradiation originating from the wind photosphere alone can explain the optical SED, and the hard irradiation is constrained to have a fraction $f_{\rm out} < 10^{-2}$. 

In our sample, there are two sources, NGC 1313 X-2 and NGC 5204 X-1, where hard irradiation may be needed in the fitting. This may suggest that they have a different geometry, in which a considerable fraction ($f_{\rm out} = 0.01 - 0.1$) of hard X-rays could reach the outer accretion disk, which could be a result of different accretion rates. However, it can also been explained as a result of uncertainties in the compact object mass or extinction, or both. For NGC 1313 X-2, a 3 solar mass neutron star can fit the data, and the fit can be significantly improved with a Galactic instead of nebula extinction. Such an extinction is possible, as \citet{Grise2011} find that the extinction in the nebula around Holmberg IX X-1 may vary by a factor up to~3.  For NGC 5204 X-1, a 15~$M_\sun$ black hole with a Galactic extinction can make up the discrepancy between the data and model.  This the highest mass allowed with X-ray spectral modeling. Thus, the inferred $f_{\rm out}$ for these two sources should be regarded as upper limits.

Disk irradiation is commonly seen in low mass X-ray binaries, and is responsible for producing the optical emission during their outbursts \citep{vanParadijs1994}. Based on the irradiation model {\tt diskir} \citep{Gierlinski2009}, the irradiation fraction $f_{\rm out}$ typically ranges from several $10^{-3}$ to a few $10^{-2}$ \citep[e.g.,][]{Chiang2010,Gandhi2010,Gierlinski2009,Kimura2019}. Although a geometry for subcritical accretion was assumed, disk irradiation in ULXs has also been studied based on the {\tt diskir} model \citep[e.g.,][]{Berghea2012,Grise2012,Kaaret2009} or the {\tt optxirr} model \citep{Sutton2014}, and the fraction of the bolometric luminosity being reprocessed on the outer disk is similar, $\sim$$10^{-3} - 10^{-2}$. We note that the definition of $f_{\rm out}$ could be different.  In {\tt diskir}, it refers to the fraction of the bolometric luminosity, while in our work, it is based on the Comptonization component in the 0.3-10 keV energy range.  Thus, by definition, our $f_{\rm out}$ is greater than that in {\tt diskir} in similar conditions, but they should be of the same order of magnitude, as the Comptonization component is the dominant spectral component in non-supersoft ULXs.  If one assumes a point-like X-ray source residing in the center of a standard accretion disk with isotropic emission, the fractional solid angle intercepted by the outer disk (e.g., from $0.1 \times r_{\rm out}$ to $r_{\rm out}$) is a few $10^{-3}$, consistent with the numbers derived with above models.  These suggest that, in the case of no beaming, the irradiation fraction $f_{\rm out}$ is expected to be $\sim$$10^{-3} - 10^{-2}$. 

Numerical simulations for supercritical accretion indicate that the emission is mildly beamed toward the system axis. As a result, the emergent flux decreases with increasing viewing angle \citep{Kawashima2012, Sadowski2015}; the inferred isotropic luminosity observed at $i \sim 80-90^\circ$ is about 1--2 orders of magnitude lower than that observed at $i \sim 0^\circ$ if the system has a mass accretion rate $\dot{m} \approx 100-300$. On the contrary,  \citet{Jiang2014} present an opposite picture, which shows that the emission is nearly isotropic, varying by a factor of less than 20\% at different viewing angles.  

As the inferred upper limit on $f_{\rm out}$ ($\sim$$10^{-2}$) in this work is higher than the irradiation fraction ($\sim$$10^{-3} - 10^{-2}$) in the case of no beaming, we cannot place a stringent constraint on the degree of beaming in ULXs, nor can we distinguish the simulation results. This is limited by the current quality of data. In the future, SEDs covering a wide wavelength band, especially in the ultraviolet band, would be helpful for this purpose.

\begin{acknowledgments}
We thank the anonymous referee for useful comments. HF acknowledges funding support from the National Key R\&D Project (grant No.\ 2018YFA0404502), the National Natural Science Foundation of China (grants Nos.\ 11633003, 12025301 \& 11821303), and the CAS Strategic Priority Program on Space Science (grant No.\ XDA15020501-02).
\end{acknowledgments}


\end{document}